\documentclass[12pt,times]{article}
 \usepackage[pdftex]{color,graphicx}
\usepackage{enumerate}
\usepackage{amsmath}
\usepackage{amssymb}
\usepackage{alltt}
\usepackage{setspace}
\usepackage{subfigure}
\usepackage{sidecap}
\usepackage{sectsty}
\usepackage{wasysym}
\partfont{\large}
\sectionfont{\large}
\subsectionfont{\small}
\usepackage[left=2cm,top=2cm,right=3cm,head=1cm,foot=1cm]{geometry}
\usepackage{mathrsfs}
\usepackage[T1]{fontenc}
\numberwithin{equation}{section}
\let\oldsqrt\sqrt
\def\sqrt{\mathpalette\DHLhksqrt}
\def\DHLhksqrt#1#2{%
\setbox0=\hbox{$#1\oldsqrt{#2\,}$}\dimen0=\ht0
\advance\dimen0-0.2\ht0
\setbox2=\hbox{\vrule height\ht0 depth -\dimen0}%
{\box0\lower0.4pt\box2}}
\newcommand{\al}{\alpha}

\newcommand{\e}{\varepsilon}
\newcommand{\ta}{\theta}

\newcommand{\ph}{\varphi}

\newcommand{\pa}{\partial}
\newcommand{\na}{\nabla}

\newcommand{\Ld}{\Lambda}

\newcommand{\ml}{\left(\begin{matrix}}
\newcommand{\mr}{\end{matrix}\right)}

\newcommand{\W}{\Omega}

\newcommand{\op}{\mathcal O}
\newcommand{\del}{\delta}
\newcommand{\Del}{\Delta}

\newcommand{\re}{\text{Re}}
\newcommand{\im}{\text{Im}}

\newcommand{\half}{\tfrac{1}{2}}

\newcommand{\s}{\sigma}

\newcommand{\Ds}{\mathscr D}

\newcommand{\As}{\mathscr A}

\newcommand{\Ms}{\mathscr M}

\newcommand{\THORN}{\text{\th}}
\newcommand{\ETH}{\text{\dh}}

\begin{document}
\title{Gravitational shockwave on the Kerr-AdS horizon}
\date{}
\author{Yoni BenTov \\ \small Perimeter Institute for Theoretical Physics, Waterloo, Ontario N2L 2Y5}
\maketitle

\begin{abstract}
	I generalize the Dray-'t Hooft gravitational shockwave to the Kerr-AdS background.  
\end{abstract}

\section{Introduction}\label{sec:intro}

Shenker and Stanford \cite{shenker_stanford}, Kitaev \cite{kitaev2018}, and Polchinski \cite{polchinski2017a} have emphasized the importance of the butterfly effect in constructing a statistical mechanics for black holes. Motivated by that, Swearngin and I \cite{bentov_swearngin} generalized the Dray-'t Hooft gravitational shockwave \cite{dray_thooft} to the Kerr-Newman spacetime. A tried and true way to regularize a physical system is to put it in a box, so a natural next step would be to consider asymptotically anti-de Sitter (AdS) boundary conditions. 
\\\\
Ask, and it shall be given: I will calculate the gravitational backreaction from a massless particle on the future horizon of an uncharged\footnote{Charge was helpful when developing the method, but I have no further use for it.} rotating black hole with negative cosmological constant. 
\\\\
I will open in Sec.~\ref{sec:background} with a review of the Kerr-AdS spacetime using the language of spin coefficients. Then in Sec.~\ref{sec:shift}, following my previous work with Swearngin, I will shift the frame to model backreaction---my main result is the generalized Dray-'t Hooft operator in Eqs.~(\ref{eq:alt result}) and~(\ref{eq:mass term}). 
\\\\
This time, however, I will also derive rotational corrections to the shockwave's angular profile. Toward doing so, in Sec.~\ref{sec:bifurcation surface} I will review the bifurcation-surface geometry for Kerr-AdS, and then in Sec.~\ref{sec:slow rotation} I will acquiesce to perturbation theory. Finally, in Sec.~\ref{sec:eom}, I will solve Einstein's equation to second order in angular momentum. I will briefly conclude in Sec.~\ref{sec:end} with a plan for where to go next. 

\section{Kerr-AdS spacetime}\label{sec:background}

My account of Kerr-AdS will be brief and utilitarian. I will assume familiarity with the exterior Kerr geometry as described by the Newman-Penrose (NP) and Geroch-Held-Penrose (GHP) formalisms in mostly-plus signature. The uninitiated should first work through the corresponding section of my previous work with Swearngin. 

\subsection{Null frame}\label{sec:frame}

I will use coordinates $(t,r,\mu,\ph)$ (with $\mu \equiv \cos\ta$), which I will call ``Schwarzschild-like'' because in the limit of zero cosmological constant they will reduce to Boyer-Lindquist coordinates for the Kerr spacetime.
\\\\
The parameters of the Kerr-AdS spacetime are $M$, $a$, and $\ell \equiv \sqrt{-3/\Ld}$, all of which have dimensions of length in standard gravitational units ($G = c = 1$). I find it convenient to define the functions
\begin{align}\label{eq:variables}
&z \equiv r+ ia\mu\;,\;\;z_0 \equiv r+ia\;,\nonumber\\
&\Sigma \equiv 1-\frac{a^2}{\ell^2}\mu^2\;,\;\; \Sigma_0 \equiv 1-\frac{a^2}{\ell^2}\;.
\end{align}
The thermodynamic energy and angular momentum are $E = \Sigma_0^{-2}M$ and $J = aE$ \cite{gibbonsperrypope}, although I will not need those in what follows. What I should note, however, is that there is an upper bound on the rotation parameter \cite{hawking_hunter_taylor-robinson}:
\begin{equation}
a \leq \ell\;.
\end{equation}
When $a \to \ell$, the rotational speed of the boundary approaches the speed of light. I will consider only $0<a<\ell$.
\\\\
Horizons and poles are zeros of the functions
\begin{equation}\label{eq:functions}
\Del \equiv |z_0|^2\left(1+\frac{r^2}{\ell^2}\right)-2Mr\;,\;\;\Phi \equiv (1\!-\!\mu^2)^{1/2}\,\Sigma^{1/2}\;.
\end{equation}
When $\ell^2 \to \infty$ (i.e., $\Ld \to 0$), those functions reduce to $\Del \to r^2+a^2-2Mr$ and $\Phi \to (1-\mu^2)^{1/2} = \sin\ta$. Because $a < \ell$, the zeros of $\Phi$ remain $\mu_\pm = \pm 1$, recovering the usual North and South Poles. 
\\\\
With negative cosmological constant, the horizon function retains its Kerr-like property of having only two real zeros, denoted again by $r_\pm$. But now they are roots of a $4^{\text{th}}$-order polynomial, so their explicit forms are cumbersome. Fortunately all I will need is the implicit relation
\begin{equation}\label{eq:M in terms of horizon}
M = \frac{|z_{0+}|^2}{2r_+} \left(1+\frac{r_+^2}{\ell^2}\right)\;,
\end{equation}
where $z_{0+} = r_+ + ia$. 
\\\\
In terms of the above functions and parameters, a collection of null 1-forms that describes the exterior Kerr-AdS spacetime is
\begin{align}\label{eq:forms}
&l = -dt+\frac{|z|^2}{\Del}\,dr + a(1\!-\!\mu^2)\frac{d\ph}{\Sigma_0}\;,\;\; l' = \frac{\Del}{2|z|^2}\left[-dt-\frac{|z|^2}{\Del}\,dr+a(1\!-\!\mu^2)\frac{d\ph}{\Sigma_0}\right]\;,\nonumber\\
&m = \frac{1}{\sqrt 2 z}\left[-\frac{|z|^2}{\Phi}\,d\mu+i\Phi\left(|z_0|^2\,\frac{d\ph}{\Sigma_0}-a\,dt\right)\right]\;.
\end{align}
The corresponding vector fields are
\begin{align}\label{eq:vectors}
&D = \frac{1}{\Del}\left(|z_0|^2\pa_t + a\Sigma_0\, \pa_\ph\right)+\pa_r\;,\;\; D' = \frac{\Del}{2|z|^2}\left[\frac{1}{\Del}\left(|z_0|^2\pa_t+a\Sigma_0\,\pa_\ph\right)-\pa_r\right]\;,\nonumber\\
&\del = \frac{1}{\sqrt 2 z}\left\{ -\Phi\,\pa_\mu + \frac{i}{\Phi}\left[\Sigma_0\, \pa_\ph + a(1\!-\!\mu^2)\, \pa_t\right]\right\}\;.
\end{align}
The frame defined by Eqs.~(\ref{eq:forms}) and~(\ref{eq:vectors}) is principal (see Sec.~\ref{sec:curvature scalars}), and its congruences are geodesic and shear-free (see Sec.~\ref{sec:spin coefficients}).  

\subsection{Spin coefficients}\label{sec:spin coefficients}

The weighted spin coefficients in the above frame are
\begin{align}\label{eq:weighted spin coefficients}
&\kappa = \s = \kappa' = \s' = 0\;,\;\;\rho = \frac{1}{z^*}\;,\;\; \rho' = -\,\frac{\Del}{2|z|^2 z^*}\;,\;\;\tau = \frac{ia\Phi}{\sqrt 2 |z|^2}\;,\;\;\tau' = \frac{ia\Phi}{\sqrt 2 (z^*)^2}\;.
\end{align}
As I asserted, the congruences are geodesic ($\kappa = \kappa' = 0$) and shear-free ($\s = \s' = 0$). They also expand ($\re(\rho) > 0$ and $\re(\rho') > 0$) and rotate ($\im(\rho) \neq 0$ and $\im(\rho') \neq 0$). The timelike counterparts to expansion and twist are
\begin{equation}
\tau\!+\!\tau'^* = \frac{-\sqrt 2 a^2\mu\Phi}{z|z|^2}\;,\;\;\tau\!-\!\tau'^* = \frac{i \sqrt 2 a r \Phi}{z|z|^2}\;.
\end{equation}
The GHP-invariant products $\rho\rho'$ and $\tau\tau'$ are proportional to the horizon function and pole function, as they should be:
\begin{equation}
\rho\rho' \propto \Del\;,\;\;\tau\tau' \propto \Phi^2\;.
\end{equation}
The Kerr-AdS black hole also has an ergosphere, whose outer boundary is the Sachs-invariant surface defined by the vanishing of 
\begin{equation}
\rho\rho' - \tau\tau' = -\,\frac{\Del-a^2\Phi^2}{2|z|^2(z^*)^3}\;.
\end{equation}
Medieval orthography \cite{ghp} requires gauge fields: 
\begin{equation}
\e = 0\;,\;\;\beta = \frac{\pa_\mu\Phi}{2\sqrt2\,z}\;,\;\; \e' = \rho'+\frac{\pa_r\Del}{4|z|^2}\;,\;\;\beta' = \tau'+\beta^*\;.
\end{equation}
A couple of noteworthy GHP-invariant relations are
\begin{align}
|\tau|^2 = |\tau'|^2\;,\;\;\ETH\tau' = \ETH'\tau\;.
\end{align}
\subsection{Curvature scalars}\label{sec:curvature scalars}
In the basis of Eqs.~(\ref{eq:forms}) and~(\ref{eq:vectors}), the free gravitational curvature is purely Coulomb:
\begin{equation}
\Psi_2 = \frac{-M}{(z^*)^3}\;,\;\;\Psi_{0,1,3,4} = 0\;.
\end{equation}
So the spacetime is Type D, and the basis is principal. The matter-induced curvature comes only from the cosmological constant:
\begin{equation}
\Pi = -\,\frac{1}{2\ell^2}\;,\;\;\Phi_{ij} = 0\;\;(i,j\,=\,0,1,2)\;.
\end{equation}
That completes a spin-coefficient description of the exterior spacetime. To deal with the horizon, I will need to pass from Schwarzschild-like to Kruskal-like coordinates, and then from my chosen frame to a smooth one. 
\subsection{Kruskal-like coordinates}\label{sec:kruskal}
My horizon-penetrating coordinates will basically follow Chandrasekhar's prescription for Kerr \cite{chand}. First I define Eddington-like coordinates ($u,v,\mu,\ph$) for approaching the hole:
\begin{equation}
u \equiv t-r_*\;,\;\;v \equiv t+r_*\;,\;\; r_* \equiv \int\frac{|z_0|^2}{\Del}\;dr\;.
\end{equation}
Then I define Kruskal-like coordinates ($U,V,\mu,\chi$) for the final descent: 
\begin{equation}
U \equiv -e^{-\al u}\;,\;\;V \equiv e^{\,\al v}\;,\;\;\chi \equiv \ph-\W\, t\;.
\end{equation}
In the frame of Sec.~\ref{sec:frame}, the surface gravity is the limiting value of the outgoing inaffinity at the North Pole of the past\footnote{Performing a GHP transformation to put the factor $\frac{\Del}{2|z|^2}$ into $l$ instead of $l'$ would shuffle the surface gravity into the future horizon.} horizon:
\begin{equation}\label{eq:surface gravity}
\al \equiv \left. 2\,\re(\e') \right|_{\mu \,=\, 1,\,r\,=\,r_+} = \left.\frac{\pa_r\Del}{2|z_0|^2}\right|_{r\,=\,r_+} = \frac{r_+}{2|z_{0+}|^2}\left(1+\frac{3r_+^2}{\ell^2}+\frac{a^2}{\ell^2}-\frac{a^2}{r_+^2}\right)\;.
\end{equation}
Combining that with Eq.~(\ref{eq:M in terms of horizon}) allows me to relate surface gravity to mass. 
\\\\
The angular velocity at the horizon is
\begin{equation}\label{eq:angular velocity}
\Omega = \frac{a \Sigma_0}{|z_{0+}|^2}\;.
\end{equation}
I still do not know how to define that from $\beta$ and $\beta'$, but I will give a coordinate prescription in the next subsection. 

\subsection{Smooth frame}\label{sec:smooth frame}
To erect a smooth frame, I will perform the following GHP transformation:
\begin{equation}\label{eq:GHP transformation}
\hat l \equiv -U\,l\;,\;\; \hat l' \equiv -U^{-1}\,l'\;,\;\;\hat m \equiv m\;.
\end{equation}
I will also define the following function to streamline the notation:
\begin{equation}\label{eq:C}
C(r) \equiv -\,\frac{\Del}{UV}\;.
\end{equation}
Applying the above procedure to the frame in Eq.~(\ref{eq:forms}), I find a basis of smooth 1-forms:
\begin{align}\label{eq:smooth forms}
&\hat l= \frac{-1}{2\al}\left[1\!+\!\frac{|z|^2}{|z_0|^2}\!-\!\frac{(1\!-\!\mu^2)a\Omega}{\Sigma_0}\right]dU + \frac{U}{2\al V}\left[1\!-\!\frac{|z|^2}{|z_0|^2}\!-\!\frac{(1\!-\!\mu^2)a\Omega}{\Sigma_0}\right]dV\!-\!U\frac{(1\!-\!\mu^2)a}{\Sigma_0}\,d\chi\;,\nonumber\\
&\hat l' = \frac{-C}{4\al |z|^2}\left[1\!+\!\frac{|z|^2}{|z_0|^2}\!-\!\frac{(1\!-\!\mu^2)a\Omega}{\Sigma_0}\right]dV+\frac{C}{4\al |z|^2}\frac{V}{U}\left[1\!-\!\frac{|z|^2}{|z_0|^2}\!-\!\frac{(1\!-\!\mu^2)a\Omega}{\Sigma_0}\right]dU\!+\! V\frac{C a (1\!-\!\mu^2)}{2|z|^2\Sigma_0}d\chi\;,\nonumber\\
&\hat m = \frac{1}{\sqrt 2 z}\left\{ -\frac{|z|^2}{\Phi}d\mu+i\Phi\left[ \frac{|z_0|^2}{\Sigma_0}d\chi+\frac{a}{2\al}\left( 1-\frac{|z_0|^2\Omega}{a\Sigma_0} \right)\left(\frac{dU}{U}-\frac{dV}{V}\right) \right] \right\}\;.
\end{align}
The angular velocity in Eq.~(\ref{eq:angular velocity}) can now be recovered by demanding $\left.\hat l\right|_{U\, =\, V\, =\, 0} \propto dU$.
\\\\
For the corresponding vector fields, I obtain:
\begin{align}\label{eq:smooth vectors}
&\hat D = \frac{2\al |z_0|^2}{C}\,\pa_V + \frac{a\Sigma_0}{CV}\left(1-\frac{|z_0|^2 \Omega}{a\Sigma_0}\right) \pa_\chi\;,\;\;\hat D' = \frac{\al |z_0|^2}{|z|^2}\,\pa_U - \frac{a\Sigma_0}{2|z|^2 U}\left(1-\frac{|z_0|^2\Omega}{a\Sigma_0}\right)\pa_\chi\;,\nonumber\\
&\hat \del = \frac{1}{\sqrt 2 z}\left\{ -\Phi\,\pa_\mu+\frac{i}{\Phi}\left[\Sigma_0-(1\!-\!\mu^2)a\Omega\right]\pa_\chi+i\frac{\al a(1\!-\!\mu^2)}{\Phi}\left(-U\pa_U+V\pa_V\right) \right\}\;.
\end{align}
My arsenal established, I enter enemy territory. 
\section{Shift}\label{sec:shift}
The first step toward dispatching backreaction is to shift the frame \cite{fels1989}:
\begin{equation}\label{eq:shift}
\tilde l \equiv l\;,\;\; \tilde l' \equiv l'+S\,l\;,\;\;\tilde m \equiv m\;.
\end{equation}
This is the Cartan perspective on generalized Kerr-Schild metrics \cite{taub1981}. In the smooth frame, my ansatz for the shift function $S$ will read
\begin{equation}\label{eq:ansatz}
\hat S \equiv (-U)^{-2} S = -\,\frac{C(r)}{2|z|^2}\,\del(U)\,f(\mu,\chi)\;.
\end{equation}
The second step comprises some preliminary identities and a merciless slog through the shifted curvature scalars. 
\\\\
Most of this amounts to a straightforward repetition of what Swearngin and I did, so I will refer you there for details. 

\subsection{Key facts}\label{sec:key facts}

Relative to the calculation without $\Ld$, some of the preliminary identities will change, while others will stay the same. Throughout what follows, it should be understood that $r$ will be set to $r_+$ by the delta function from the ansatz in Eq.~(\ref{eq:ansatz}). 
\\\\
By explicit calculation, I find that two of what Swearngin and I called ``key facts'' remain unchanged:
\begin{equation}\label{eq:old key facts}
\ETH F(r)  = 0\;,\;\; \ETH\left(\frac{1}{|z|^2}\right) = -\,\frac{1}{|z|^2}\,(\tau\!+\!\tau'^*)\;.
\end{equation}
The third key fact, however, gets modified slightly because of the additional factor in the pole function [recall Eq.~(\ref{eq:functions})]:
\begin{equation}\label{eq:modified key fact}
\ETH \del(U) = -\,\frac{\al |z|^2}{2r\Sigma}(\tau\!-\!\tau'^*)\, U\pa_U\del(U)\;.
\end{equation}
That factor further necessitates a fourth key fact:
\begin{equation}\label{eq:new key fact}
\ETH \left( \frac{1}{\Sigma} \right) = \frac{|z|^2}{\ell^2 \Sigma^2}\,(\tau\!+\!\tau'^*)\;.
\end{equation}
Everything else will track the calculation for Kerr-Newman. Game on. 
\subsection{Outgoing transverse wave}\label{sec:Psi4}
The shift from Eq.~(\ref{eq:shift}) formally induces an outgoing transverse gravitational wave: 
\begin{equation}
\tilde\Psi_4^* = \ETH\ETH S + 2\tau \ETH S\;.
\end{equation}
Passing to the smooth basis, inserting Eq.~(\ref{eq:ansatz}), and downing a red eye, I find
\begin{equation}
\hat{\tilde\Psi}_4^* = -\,\frac{C(r)}{2|z|^2}\del(U)\,\left[\ETH\ETH f + (k_1\,\tau+ k_2\,\tau'^*)\ETH f + (k_3\,\tau^2+k_4\,\tau'^{*2} + k_5\,\tau\tau'^*)f\right]\;,
\end{equation}
with
\begin{align}
&k_1 = \frac{\al |z|^2}{r\Sigma}\;,\;\; k_2 = -2\left(1+\frac{\al |z|^2}{2r\Sigma}\right)\;,\;\; k_3 = \frac{\al |z|^4}{2r\Sigma^2}\left(\frac{\al}{2r}+\frac{1}{\ell^2}\right)\;,\nonumber\\
&k_4 = 2+\frac{\al |z|^2}{r\Sigma}\left[1+\frac{|z|^2}{2\Sigma}\left( \frac{\al}{2r} - \frac{1}{\ell^2} \right) \right]\;,\;\; k_5 = -\,\frac{\al |z|^2}{r\Sigma}\left( 1+\frac{\al |z|^2}{2r\Sigma}\right)\;.
\end{align}
When $\ell^2 \to \infty$, $\Sigma \to 1$, and the $k_i$ correctly reduce to those for Kerr-Newman. Alternatively, for $a \to 0$ with $\ell^2$ finite, I arrive at $\Psi_4$ for the Schwarzschild-AdS shockwave. To my knowledge, that itself is a new result.  
\subsection{Backreaction from massless particle}\label{sec:Phi22}
And now, every shockwave enthusiast's favorite:
\begin{align}
\tilde\Phi_{22} &= \re\left(\rho\THORN'\!-\!\rho'\THORN\right)S + \half\left(\ETH\ETH'+\ETH'\ETH + 2\tau \ETH'+2\tau^*\ETH\right)S \nonumber\\
&+ \left[(\rho\!-\!\rho^*)(\rho'\!-\!\rho'^*)+\ETH'\tau+\ETH\tau^*+2|\tau|^2+2\,\re(\Psi_2\!+\!2\Pi)\right]S\;.
\end{align}
Turning the cranks, I eventually mill the GHP-covariant form of the result:
\begin{align}
\hat{\tilde\Phi}_{22} &= -\,\frac{C(r)}{4|z|^2}\,\del(U)\, \Ds f\;,
\end{align}
with the differential operator
\begin{align}\label{eq:result}
\Ds &= \ETH\ETH'+\ETH'\ETH + \left\{ \left[-(\tau\!+\!\tau'^*) + \left( 1+\frac{\al |z|^2}{r\Sigma} \right)(\tau\!-\!\tau'^*) \right]\ETH' + c.c. \right\} \nonumber\\
&-(\ETH'\tau+c.c.)+\half|\tau\!+\!\tau'^*|^2+\left(1+\frac{\al |z|^2}{r\Sigma}\right)^2 \half|\tau\!-\!\tau'^*|^2 + 2[\re(\Psi_2)\!+\!2\Pi]\;.
\end{align}
Mission accomplished. 
\subsection{Mass term}\label{sec:mass term}

I will now trade the GHP derivatives for a bifurcation-surface Laplacian, hoping to excavate a mass term that goes to zero in the extremal limit \cite{sfetsos, maldacena2016b}.   
\\\\
Using $\beta' = \tau'+\beta^*$ along with $\ETH\tau'=\ETH'\tau$, I extract a Laplacian by acting on the horizon field $f(\mu,\chi) \sim (-1,-1)$: 
\begin{equation}
\ETH\ETH'+\ETH'\ETH = \na_{\!\text{2d}}^{\;2} + 4(\tau'\del + c.c.) + 2\left[(\ETH'\tau+c.c.) + |\tau\!+\!\tau'^*|^2 + |\tau\!-\!\tau'^*|^2\right]\;.
\end{equation}
This lets me put Eq.~(\ref{eq:result}) into the form
\begin{align}\label{eq:alt result}
\Ds &= \na_{\!\text{2d}}^{\;2} + \left\{ \left[(\tau\!+\!\tau'^*)+\left(-1+\frac{\al |z|^2}{r\Sigma}\right)(\tau\!-\!\tau'^*)\right]\del' + c.c. \right\}+\Ms\;,
\end{align}
with mass term
\begin{equation}\label{eq:intermediate mass term}
\Ms = (\ETH'\tau+c.c.)+\half|\tau\!+\!\tau'^*|^2+\left(1-\frac{\al |z|^2}{r\Sigma}\right)^2 \half|\tau\!-\!\tau'^*|^2+2\left[\re(\Psi_2)\!+\!2\Pi\right]\;.
\end{equation}
This intermediate form deserves a brief remark: Although I have explicitly broken GHP covariance, I recover a GHP-covariant (further, invariant) form for the mass term. 
\\\\
Regardless, the important physics follows from the dependence on surface gravity, and that dependence is not yet evident. I was originally guided toward the form of Eq.~(\ref{eq:intermediate mass term}) by comparison with the nonrotating limit, but now I will backtrack and reuse one of the GHP equations in the other direction:
\begin{equation}
\ETH'\tau + |\tau|^2 + \Psi_2 + 2\Pi = \THORN'\rho + \rho'^*\rho\;.
\end{equation}
Using that along with $2|\tau|^2 = \half|\tau\!+\!\tau'^*|^2+\half|\tau\!-\!\tau'^*|^2$, and recalling that $\rho'^*\rho = \hat\rho'^*\hat\rho = 0$ at $U = 0$, I can trade curvatures and $\tau$s for $\THORN'\rho$:
\begin{equation}
\Ms = 2\,\re(\THORN'\rho)-\frac{\al |z|^2}{r\Sigma}\left(1-\frac{\al |z|^2}{2r\Sigma}\right)|\tau\!-\!\tau'^*|^2\;.
\end{equation}
Finally, with 
\begin{equation}
\hat D'\hat\rho = -\al\,\frac{|z_0|^2}{|z|^2 z^*}\;,
\end{equation}
I get
\begin{equation}\label{eq:mass term}
\Ms = -\al\left[ \frac{2r |z_0|^2}{|z|^4}+\frac{|z|^2}{r\Sigma}\left(1-\frac{\al |z|^2}{2r\Sigma}\right) |\tau\!-\!\tau'^*|^2\right]\;.
\end{equation}
The mass term is proportional to surface gravity and therefore vanishes in the extremal limit. Victory.
\\\\
In the last paper, having developed the entire computational apparatus from scratch, Swearngin and I rightly ended there---with Eqs.~(\ref{eq:result}), ~(\ref{eq:alt result}), and~(\ref{eq:mass term}) behind me, I too could wrap up and live happily ever after. 
\\\\
Instead, I will calculate rotational corrections to the angular profile. Buckle up. 


\section{Bifurcation-surface geometry}\label{sec:bifurcation surface}

To clear the on-ramp, I need the Laplacian. From the smooth 1-forms in Eq.~(\ref{eq:smooth forms}), I find
\begin{equation}
\left.\hat m\right|_{U\,=\,V\,=\,0} = \frac{1}{\sqrt 2 z} \left(-\,\frac{|z|^2}{\Phi}\,d\mu + i\Phi\,\frac{|z_0|^2}{\Sigma_0}\,d\chi\right)\qquad(r\,=\,r_+)\;.
\end{equation}
The line element for the surface is then (recall $\Phi^2 = (1\!-\!\mu^2)\,\Sigma$ with $\Sigma = 1\!-\!\frac{a^2}{\ell^2}\mu^2$)
\begin{equation}
d\s^2 \equiv 2\hat m\hat m' = \frac{|z|^2}{\Sigma}\,\frac{d\mu^2}{1\!-\!\mu^2} + \frac{\Sigma}{\Sigma_0^{\,2}}\,\frac{|z_0|^4}{|z|^2}\,(1\!-\!\mu^2)\,d\chi^2\;.
\end{equation}
When $\ell^2 \to \infty$, $\Sigma$ and $\Sigma_0$ reduce to 1, and the line element becomes that of a squashed sphere. When $a \to 0$ for fixed finite $\ell^2$, $\Sigma$ and $\Sigma_0$ again become 1, but this time $z$ and $z_0$ further reduce to $r$, and the line element becomes that of an ordinary sphere. 
\\\\
At any rate, the Laplacian derived from $d\s^2$ is
\begin{equation}\label{eq:2d laplacian}
\na_{\!\text{2d}}^{\,2} = \frac{\Sigma}{|z|^2}\left\{ \pa_\mu\left[(1\!-\!\mu^2)\pa_\mu\right]-\frac{2\mu(1\!-\!\mu^2)a^2}{|z|^2}\left(1+\frac{|z|^2}{\ell^2\Sigma}\right)\pa_\mu+\frac{1}{1\!-\!\mu^2}\left(\frac{\Sigma_0}{\Sigma}\frac{|z|^2}{|z_0|^2}\right)^2\pa_\chi^{\,2} \right\}\;.
\end{equation}
When $a = 0$, this reduces to 
\begin{equation}
\left. \na_{\!\text{2d}}^{\,2} \right|_{a\,=\,0} = \frac{1}{r^2}\na_{\!\text{o}}^{\,2}\;,\;\; \na_{\!\text{o}}^{\,2} = \pa_\mu\left[(1\!-\!\mu^2)\pa_\mu\right]+\frac{1}{1\!-\!\mu^2}\,\pa_\chi^{\,2}\;.
\end{equation}
I write that only to emphasize the relation of Eq.~(\ref{eq:2d laplacian}) to the typical Laplacian on the \textit{unit} sphere, $\na_{\!\text{o}}^{\,2}$. For the rotating case, the functions $|z|^2 = r^2 + a^2\mu^2$ and $\Sigma = 1-\frac{a^2}{\ell^2}\mu^2$ depend on $\mu = \cos\ta$, and I must proceed with caution. 

\subsection{Expand in spherical harmonics}

I am not a conjurer of special functions, so I will stick to what I learned in school: Spherical harmonics. 
\\\\
Let $P_{lm}(\mu)$ denote the associated Legendre function that diagonalizes the longitudinal part of the spherical Laplacian:
\begin{equation}\label{eq:associated legendre}
\pa_\mu\left[(1\!-\!\mu^2\right)\pa_\mu P_{lm}(\mu)] = \left(\frac{m^2}{1\!-\!\mu^2} - l(l+1)\right)\,P_{lm}(\mu)\;.
\end{equation}
I will expand the shockwave's angular profile as follows:
\begin{equation}\label{eq:expand in spherical harmonics}
f(\mu,\chi) \equiv \sum_{l\,=\,0}^\infty \sum_{m\,=\,-l}^l\, f_{lm}\,P_{lm}(\mu)\,e^{\,im\chi}\;.
\end{equation}
The Laplacian's first and third terms are taken care of by Eq.~(\ref{eq:associated legendre}):
\begin{align}
&\left\{\pa_\mu\left[(1\!-\!\mu^2)\pa_\mu\right]+\frac{1}{1\!-\!\mu^2}\left(\frac{\Sigma_0}{\Sigma}\frac{|z|^2}{|z_0|^2}\right)^2 \pa_\chi^{\,2}\right\} \left[P_{lm}(\mu)\,e^{\,im\chi}\right] \nonumber\\
&\qquad= \left\{-l(l\!+\!1)+\left[1-\left(\frac{\Sigma_0}{\Sigma}\frac{|z|^2}{|z_0|^2}\right)^2\right]\frac{m^2}{1\!-\!\mu^2}\right\}\,P_{lm}(\mu)\,e^{\,im\chi}\;. 
\end{align}
For its middle term, I will require two standard recurrence relations:
\begin{align}
&\mu\,P_{lm} = \frac{1}{2l\!+\!1}\left[(l\!+\!m)P_{l-1,m}+(l\!-\!m\!+\!1)P_{l+1,m}\right]\;, \label{eq:mu Plm} \\
&\pa_\mu P_{lm} = \frac{1}{1\!-\!\mu^2}\left[(l\!+\!1)\,\mu P_{lm}-(l\!-\!m\!+\!1)P_{l+1,m}\right]\;. \label{eq: d Plm}
\end{align}
Inserting the first of those into the second gives the useful relation
\begin{equation}\label{eq:derivative of Plm}
(1\!-\!\mu^2)\, \pa_\mu P_{lm} = \frac{1}{2l\!+\!1}\left[(l\!+\!1)(l\!+\!m)P_{l-1,m}-l(l\!-\!m\!+\!1)P_{l+1,m}\right]\;.
\end{equation}
Multiplying by $\mu$ and using Eq.~(\ref{eq:mu Plm}), I deduce the needed formula:
\begin{align}
\mu(1\!-\!\mu^2)\,\pa_\mu P_{lm} &= \frac{1}{2l\!+\!1}\left\{  \frac{(l\!+\!1)(l\!+\!m)(l\!+\!m\!-\!1)}{2l\!-\!1}\,P_{l-2,m} - \frac{l(l\!-\!m\!+\!1)(l\!-\!m\!+\!2)}{2l\!+\!3}\,P_{l+2,m}  \right. \nonumber\\
& \left. + \left[\frac{(l\!+\!1)(l\!+\!m)(l\!-\!m)}{2l\!-\!1}-\frac{l(l\!-\!m\!+\!1)(l\!+\!m\!+\!1)}{2l\!+\!3}\right] P_{lm}\right\}\;.
\end{align}
The Kerr-AdS bifurcation-surface Laplacian acting on $P_{lm}(\mu)\,e^{\,im\chi}$ is then
\begin{align}\label{eq:2d laplacian on f}
&\na_{\!\text{2d}}^{\,2}(P_{lm}\,e^{\,im\chi}) = \frac{\Sigma}{|z|^2}\,e^{\,im\chi}\left\{ -\,\frac{2a^2(1+\frac{|z|^2}{\ell^2\Sigma})}{|z|^2(2l\!+\!1)} \left[\tfrac{(l+1)(l+m)(l+m-1)}{2l-1}\,P_{l-2,m} - \tfrac{l(l-m+1)(l-m+2)}{2l+3}\,P_{l+2,m} \right]  \right. \nonumber\\
&\left. +\left[-l(l\!+\!1)+\left(1\!-\!\left(\tfrac{\Sigma_0}{\Sigma}\tfrac{|z|^2}{|z_0|^2}\right)^2\right)\frac{m^2}{1\!-\!\mu^2} -\frac{2a^2 (1\!+\!\frac{|z|^2}{\ell^2\Sigma})}{|z|^2(2l\!+\!1)}\left( \tfrac{(l+1)(l+m)(l-m)}{2l-1}-\tfrac{l(l-m+1)(l+m+1)}{2l+3} \right)\right]P_{lm} \right\}\;.
\end{align}
Since the surface is deformed relative to Schwarzschild's by $a$ and $\ell^2$, the spherical Laplacian mixes modes of different $l$. But because even the deformed surface remains axisymmetric, modes of different $m$ do not mix. 
\subsection{Single NP derivatives}

The rotating Dray-'t~Hooft operator in Eq.~(\ref{eq:alt result}) contains single derivatives beyond those from the 2d Laplacian. So I will return to Eq.~(\ref{eq:smooth vectors}) for an angular vector field:
\begin{equation}
\left.\hat\del \right|_{U\,=\,V\,=\,0} = \frac{1}{\sqrt2 z}\left(-\Phi\,\pa_\mu+i\,\frac{\Sigma_0}{\Phi}\,\frac{|z|^2}{|z_0|^2} \pa_\chi\right)\qquad(r = r_+)\;.
\end{equation}
Acting on $P_{lm}(\mu)\,e^{\,im\chi}$ and using the recurrence relations, I find
\begin{equation}\label{eq:delta f}
\hat \del (P_{lm}(\mu)\,e^{\,im\chi}) = \frac{-\Sigma}{\sqrt2 z \Phi}\,e^{\,im\chi}\left\{ \frac{1}{2l\!+\!1}\left[(l\!+\!1)(l\!+\!m)\,P_{l-1,m} - l(l\!-\!m\!+\!1)\,P_{l+1,m} \right] + m\,\frac{\Sigma_0}{\Sigma} \frac{|z|^2}{|z_0|^2}\,P_{lm} \right\}\;.
\end{equation}
Because both $\del'$ and $e^{im\chi}$ are complex, it is worth writing $\del'(P_{lm}e^{\,im\chi})$ explicitly:
\begin{equation}\label{eq:delta' f}
\hat\del'\left(P_{lm}(\mu)\,e^{\,im\chi}\right) = \frac{-\Sigma}{\sqrt2 z^*\Phi}\,e^{\,im\chi}\left\{ \frac{1}{2l\!+\!1}\left[(l\!+\!1)(l\!+\!m)\,P_{l-1,m} - l(l\!-\!m\!+\!1)\,P_{l+1,m} \right] - m\,\frac{\Sigma_0}{\Sigma} \frac{|z|^2}{|z_0|^2}\,P_{lm} \right\}\;.
\end{equation}
Note the relative signs of the $m P_{lm}$ terms in Eqs.~(\ref{eq:delta f}) and~(\ref{eq:delta' f}). 
\\\\
The single-derivative operator of interest is (recall that $\tau = -\frac{z}{z^*}\tau'^*$)
\begin{equation}\label{eq:O}
\op \equiv \left[(\tau\!+\!\tau'^*)+\left(-1+\frac{\al |z|^2}{r\Sigma}\right)(\tau\!-\!\tau'^*)\right]\del'+c.c. = 2\tau'\left(1-\frac{\al z^*}{\Sigma}\right)\del+c.c.
\end{equation}
Using Eq.~(\ref{eq:delta f}) and $\tau' = \frac{ia\Phi}{\sqrt2 (z^*)^2}$, I find the preliminary expression
\begin{equation}
\op f = \frac{-ia}{|z|^2}\left(\frac{\Sigma}{z^*}-\al \right) \As + \frac{ia}{|z|^2}\left(\frac{\Sigma}{z}-\al\right)\bar\As\;,
\end{equation}
with
\begin{equation}\label{eq:A}
\As = \sum_{l,m} e^{im\chi} f_{lm} \left\{\frac{1}{2l\!+\!1}\left[(l\!+\!1)(l\!+\!m) P_{l-1,m} - l(l\!-\!m\!+\!1) P_{l+1,m}\right]+m\frac{\Sigma_0}{\Sigma}\frac{|z|^2}{|z_0|^2} P_{lm} \right\}\;,
\end{equation}
and
\begin{equation}\label{eq:Abar}
\bar\As = \sum_{l,m} e^{-im\chi} f_{lm}^* \left\{\frac{1}{2l\!+\!1}\left[(l\!+\!1)(l\!+\!m) P_{l-1,m} - l(l\!-\!m\!+\!1) P_{l+1,m}\right]-m\frac{\Sigma_0}{\Sigma}\frac{|z|^2}{|z_0|^2} P_{lm} \right\}\;.
\end{equation}
Since the horizon field $f(\mu,\chi)$ is real, and since the associated Legendre polynomials satisfy
\begin{equation}
P_{l,-m} = (-1)^m \frac{(l\!-\!m)!}{(l\!+\!m)!} P_{lm}\;,
\end{equation}
the coefficients in Eq.~(\ref{eq:expand in spherical harmonics}) obey a reality condition:
\begin{equation}
f_{lm}^{\,*} = (-1)^m\,\frac{(l\!-\!m)!}{(l\!+\!m)!}\,f_{l,-m}\;.
\end{equation}
With those, Eq.~(\ref{eq:Abar}) can be rewritten as
\begin{equation}\label{eq:better Abar}
\bar\As = \sum_{l,m} e^{\,im\chi} f_{lm} \left\{ \frac{1}{2l\!+\!1}\left[(l\!+\!1)(l\!+\!m)\,P_{l-1,m} - l(l\!-\!m\!+\!1)\, P_{l+1,m} \right] + m\, \frac{\Sigma_0}{\Sigma} \frac{|z|^2}{|z_0|^2}\, P_{lm} \right\}\;.
\end{equation}
So the single-derivative operator in Eq.~(\ref{eq:O}) acting on the horizon field is
\begin{align}\label{eq:op f prelim}
\op f &= \sum_{l,m}\frac{2a^2\Sigma}{|z|^2}\,e^{\,im\chi} f_{lm} \left\{ \frac{1}{(2l+1)|z|^2}\left[(l\!+\!1)(l\!+\!m)\,\mu P_{l-1,m} - l(l\!-\!m\!+\!1)\,\mu P_{l+1,m}\right]+\frac{m}{|z_0|^2}\frac{\Sigma_0}{\Sigma}\,\mu P_{lm} \right\}\;.
\end{align}
Given the recurrence relation in Eq.~(\ref{eq:mu Plm}), I can further simplify Eq.~(\ref{eq:op f prelim}):
\begin{align}\label{eq:op f}
\op f &= \sum_{l,m} \frac{2a^2\Sigma}{(2l\!+\!1)|z|^4}\,e^{\,im\chi} f_{lm} \left\{  \tfrac{(l+1)(l+m)(l+m-1)}{2l-1}\,P_{l-2,m} - \tfrac{l(l-m+1)(l-m+2)}{2l+3}\,P_{l+2,m} \right. \nonumber\\
&\left. +m\frac{|z|^2}{|z_0|^2}\frac{\Sigma_0}{\Sigma}\left[(l\!+\!m)\,P_{l-1,m} + (l\!-\!m\!+\!1)\,P_{l+1,m} \right] + \left[ \tfrac{(l+1)(l+m)(l-m)}{2l-1} - \tfrac{l(l-m+1)(l+m+1)}{2l+3} \right] P_{lm}   \right\}\;.
\end{align}
That will do. 
\subsection{Dray-'t Hooft operator}
Swerving past Eqs.~(\ref{eq:2d laplacian on f}) and~(\ref{eq:op f}), I barrel toward a series representation for the generalized Dray-'t Hooft derivative of the angular profile:

\begin{align}\label{eq:series result}
\Ds f &= \sum_{l,m} e^{\,im\chi} f_{lm}\,\frac{\Sigma}{|z|^2}\left\{ k_{lm}\, P_{lm} + \frac{2a^2}{(2l\!+\!1)\Sigma}\left[ \frac{m\Sigma_0}{|z_0|^2}\left( (l\!+\!m) \,P_{l-1,m} \phantom{\frac{a}{b}}\!\!\!\!\!+ (l\!-\!m\!+\!1) \,P_{l+1,m} \right) \right. \right. \nonumber\\
&\left.\left.+ \frac{1}{\ell^2}\left( -\,\tfrac{(l+1)(l+m)(l+m-1)}{2l-1} P_{l-2,m} + \tfrac{l(l-m+1)(l-m+2)}{2l+3} P_{l+2,m} \right) \right] \right\}\;,
\end{align}
where
\begin{equation}\label{eq:klm}
k_{lm} = -l(l+1)+\left[1-\left(\tfrac{\Sigma_0}{\Sigma}\tfrac{|z|^2}{|z_0|^2}\right)^2\right] \frac{m^2}{1\!-\!\mu^2} + \frac{|z|^2}{\Sigma}\Ms - \frac{2a^2}{(2l\!+\!1)\ell^2\Sigma}\left[ \tfrac{(l+1)(l+m)(l-m)}{2l-1} - \tfrac{l(l-m+1)(l+m+1)}{2l+3} \right] \;.
\end{equation}
At this mile marker, everything remains exact. 

\section{Slow-rotation approximation}\label{sec:slow rotation}

Maybe someone more sophisticated could make sense of Eq.~(\ref{eq:series result}) as written, but I will content myself with an $O(a^2)$ approximation. 
\\\\
Since the $P_{l\pm1,m}$ and $P_{l\pm2,m}$ terms in Eq.~(\ref{eq:series result}) are already multiplied by $a^2$, the only real work to do is in approximating the function $k_{lm}$ in Eq.~(\ref{eq:klm}). First, I need
\begin{align}
&1-\left(\frac{\Sigma_0}{\Sigma}\frac{|z|^2}{|z_0|^2}\right)^2 \approx \frac{2a^2}{r^2}(1\!-\!\mu^2)\left(1+\frac{r^2}{\ell^2}\right)\;.
\end{align}
That factor of $1-\mu^2$ conveniently cancels the $\frac{1}{1-\mu^2}$ in Eq.~(\ref{eq:klm}). Next, I must expand the mass term:
\begin{equation}
\Ms \approx -\,\frac{2\al}{r}\left\{1+\left[\left(2-\frac{\al r}{2}\right)+\left(-3+\frac{\al r}{2}\right)\mu^2\right]\frac{a^2}{r^2}\right\}\;.
\end{equation}
Important stylistic remark: Because $\al$ does not depend on $\mu$, I find it convenient to leave $\al$ as is for now, with the understanding that eventually it too must be expanded around its nonrotating value. The same holds for $r$ itself---I will get to that when the time is right.  
\\\\
Recalling the additional factor multiplying $\Ms$, 
\begin{equation}
\frac{|z|^2}{\Sigma} \approx \left[1+\frac{a^2}{r^2}\left(1+\frac{r^2}{\ell^2}\right)\mu^2\right] r^2\;,
\end{equation}
I get
\begin{equation}
\frac{|z|^2}{\Sigma}\Ms \approx -2\al r\left\{ 1 + \left[ 2 - \frac{\al r}{2} + \left(-2 + \frac{\al r}{2} + \frac{r^2}{\ell^2}\right) \mu^2 \right] \frac{a^2}{r^2}\right\}\;.
\end{equation}
Why show this elementary work? Because it obnoxiously begot a $\mu^2$ that I will have to send to recurrence daycare: 
\begin{align}\label{eq:daycare}
\mu^2 P_{lm} &= \frac{1}{2l\!+\!1}\left[(l\!+\!m)\,\mu P_{l-1,m} + (l\!-\!m\!+\!1)\,\mu P_{l+1,m} \right] \nonumber\\
&= \frac{1}{2l\!+\!1}\left[ \tfrac{(l+m)(l+m-1)}{2l-1}\,P_{l-2,m} + \tfrac{(l-m+1)(l-m+2)}{2l+3}\,P_{l+2,m} + \left( \tfrac{(l+m)(l-m)}{2l-1} + \tfrac{(l-m+1)(l+m+1)}{2l+3} \right) P_{lm} \right]\;.
\end{align}
%
%
With that, all details of remote conceptual intricacy have been explained. Behold the approximate shifted $\Phi_{22}$:
\begin{equation}\label{eq:approximate series result 2}
\hat{\tilde\Phi}_{22} \approx \frac{-C}{4|z|^2}\,\del(U)\sum_{l,m} f_{lm}\,e^{\,im\chi}\frac{\Sigma}{|z|^2} \left[ u_{lm} \,P_{lm}+\frac{2a^2}{r^2}\left( v_{lm} \,P_{l+1,m} + w_{lm} \,P_{l-1,m} + x_{lm} \,P_{l+2,m} + y_{lm} \,P_{l-2,m} \right) \right]\;,
\end{equation}
where
\begin{align}
&u_{lm} = -\left[ 2\al r+l(l\!+\!1)\right] + \frac{2a^2}{r^2}\left\{ (1\!+\!\tfrac{r^2}{\ell^2})m^2-(2\!-\!\tfrac{\al r}{2}) \al r \right. \nonumber\\
&\left. +\tfrac{(l+m)(l-m)}{(2l+1)(2l-1)}\left[ (2\!-\!\tfrac{\al r}{2})\al r-(l\!+\!1\!+\!\al r)\tfrac{r^2}{\ell^2} \right] +\tfrac{(l-m+1)(l+m+1)}{(2l+1)(2l+3)}\left[ (2\!-\!\tfrac{\al r}{2})\al r + (l\!-\!\al r)\tfrac{r^2}{\ell^2} \right]\right\}\;, \label{eq:u_lm}\\
&\nonumber\\
&v_{lm} = \tfrac{m(l-m+1)}{2l+1}\;,\;\; w_{lm} = \tfrac{m(l+m)}{2l+1}\;,\;\; x_{lm} = \tfrac{(l-m+1)(l-m+2)}{(2l+1)(2l+3)}\left[ (2\!-\!\tfrac{\al r}{2}) \al r + (l\!-\!\al r)\tfrac{r^2}{\ell^2} \right]\;,\nonumber\\
&y_{lm} = \tfrac{(l+m)(l+m-1)}{(2l+1)(2l-1)}\left[(2\!-\!\tfrac{\al r}{2})\al r - (l\!+\!1\!+\!\al r)\tfrac{r^2}{\ell^2}\right]\;.
\end{align}
Warning: The superficially leading-order part of $u_{lm}$ contains both $O(a^0)$ and $O(a^2)$ parts. But the coefficients $v_{lm}, w_{lm}, x_{lm}$, and $y_{lm}$ are already multiplied by $a^2$ in Eq.~(\ref{eq:approximate series result 2}), so in those expressions both $r = r_+$ and $\al$ may be safely replaced by their nonrotating counterparts. 

\section{Equation of motion}\label{sec:eom}

Time to step on it. The Einstein equation requires 
\begin{equation}\label{eq:eom}
\hat{\tilde\Phi}_{22} = \hat{\tilde t}_{22}\;,
\end{equation} 
with the pertinent energy scalar describing a massless point-source on the future horizon:
\begin{equation}
\hat{\tilde t}_{22} = E\,\del(U)\,\del^2(\vec x\!-\!\vec x_N)\;.
\end{equation}
Here $E$ is a constant proportional to the energy of the source, $\vec x$ describes a general point on the bifurcation surface, and $\vec x_N$ points to the North Pole. In the coordinates I have been using, the 2d delta function reads
\begin{equation}\label{eq:angular delta function}
\del^2(\vec x \!-\! \vec x_N) = \frac{\Sigma_0}{|z_0|^2}\,\del(\mu\!-\!1)\,\del(\chi)\;.
\end{equation}
Back in Eq.~(\ref{eq:approximate series result 2}), I had factored out an overall $\frac{\Sigma}{|z|^2}$ to anticipate dividing Eq.~(\ref{eq:eom}) by it and enlisting Eq.~(\ref{eq:angular delta function}) to set $\mu = 1$. Proceeding along those lines, integrating the whole equation by $\frac{1}{2\pi} \int_0^{2\pi}\!d\chi\, e^{-im\chi}$, and defining the constant 
\begin{equation} \label{eq:q}
q \equiv \frac{2E|z_0|^2}{\pi C}\;,
\end{equation}
I arrive at the following form of Eq.~(\ref{eq:eom}):
\begin{equation}\label{eq:eom series}
\sum_l f_{lm}\left[u_{lm}\,P_{lm} + \frac{2a^2}{r^2} \left(v_{lm}\,P_{l+1,m} + w_{lm}\,P_{l-1,m} + x_{lm}\,P_{l+2,m} + y_{lm}\,P_{l-2,m} \right) \right] = -q\,\del(\mu\!-\!1)\;.
\end{equation}
Because all of the coefficients in this expression are \textit{constant}, I can now use the orthogonality of associated Legendre polynomials:
\begin{equation}
\int_{-1}^1 d\mu\,P_{lm}(\mu)\,P_{l'm}(\mu) = \frac{2(l\!+\!m)!}{(2l\!+\!1)(l\!-\!m)!}\,\del_{ll'}\;.
\end{equation}
Carrying this out term by term, I obtain the algebraic relation
\begin{equation}\label{eq:algebraic eom}
u_{lm}\,f_{lm} + \frac{2a^2}{r^2}\left(v_{l-1,m}\,f_{l-1,m}+w_{l+1,m}\,f_{l+1,m} + x_{l-2,m}\,f_{l-2,m} + y_{l+2,m}\,f_{l+2,m}\right) = -(l+\half)\,q\,\del_{m0}\;.
\end{equation}
The finish line beckons: Separate the lingering factors of $a^2$. 
\subsection{Approximate horizon and surface gravity}
I will parameterize the horizon's Schwarzschild-like coordinate location as
\begin{equation}\label{eq:approximate horizon}
r \equiv r_{(0)} + \frac{a^2}{r_{(0)}^2}\,r_{(2)} + O(a^4)\; \implies r^2 = r_{(0)}^2 + \frac{2a^2}{r_{(0)}}\,r_{(2)} + O(a^4)\;.
\end{equation}
Inserting that into Eq.~(\ref{eq:M in terms of horizon}) and matching powers of $a^2$ recovers the standard zeroth-order expression and further implies
\begin{equation}
r_{(2)} = \frac{-M}{\frac{M}{r_{(0)}} + \frac{2r_{(0)}^2}{\ell^2}}\;.
\end{equation}
When $\ell^2 \to \infty$, this reproduces the Kerr relation $r = M+(M^2-a^2)^{1/2} = 2M-\frac{a^2}{2M} + O(a^4)$.
\\\\
Next I will insert Eq.~(\ref{eq:approximate horizon}) into Eq.~(\ref{eq:surface gravity}) to obtain an approximation for the surface gravity. Actually, recalling $u_{lm}$ from Eq.~(\ref{eq:u_lm}), I will need only the combination 
\begin{equation}\label{eq:approximate surface gravity}
2\al r = 2\al_{(0)} r_{(0)} - \frac{2a^2}{r_{(0)}^2}\left( 1 + \frac{r_{(0)}^2}{\ell^2} - \frac{3r_{(0)}}{\ell^2}\,r_{(2)} \right) + O(a^4)\;.
\end{equation}
Instances of $\al$ that appear elsewhere are already multiplied by $a^2$ and can therefore be replaced by $\al_{(0)}$. 
\\\\
Strictly speaking, the last geometrical quantity I will need to approximate is $C \equiv -\frac{\Del}{UV}$ at $r = r_+$, because that appears in the denominator of the source term $q$ defined in Eq.~(\ref{eq:q}). This, however, would be truly beyond the pale, and I will put my foot down. The following general expression will suffice: 
\begin{equation}\label{eq:approximate q}
q \equiv q_{(0)} + \frac{2a^2}{r_{(0)}^2}\,q_{(2)} + O(a^4)\;.
\end{equation} 
\subsection{Leading rotational corrections to angular profile}
I will attempt a perturbative ansatz
\begin{equation}
f_{lm} = f_{lm}^{(0)} + \frac{2a^2}{r_{(0)}^2}\,f_{lm}^{(2)} + O(a^4)
\end{equation}
and use Eq.~(\ref{eq:approximate surface gravity}) to separate the $O(a^0)$ and $O(a^2)$ terms in $u_{lm}$ from Eq.~(\ref{eq:u_lm}):
\begin{align}
&u_{lm} = u_{lm}^{(0)} + \frac{2a^2}{r_{(0)}^2}\,u_{lm}^{(2)}\;,\;\; u_{lm}^{(0)} = -[2\al_{(0)} r_{(0)}+l(l+1)]\;, \\
&\nonumber\\
&u_{lm}^{(2)} = (1\!+\!\tfrac{r_{(0)}^2}{\ell^2})(1\!+\!m^2) - \tfrac{3r_{(0)}}{\ell^2}r_{(2)} - \al_{(0)}r_{(0)}(2\!-\!\tfrac{\al_{(0)}r_{(0)}}{2}) \nonumber\\
&\phantom{u_{lm}^{(2)}}+\tfrac{(l+m)(l-m)}{(2l+1)(2l-1)}\left[(2\!-\!\tfrac{\al_{(0)}r_{(0)}}{2}) \al_{(0)}r_{(0)}-(l\!+\!1\!+\!\al_{(0)}r_{(0)})\tfrac{r_{(0)}^2}{\ell^2}\right] \nonumber\\
&\phantom{u_{lm}^{(2)}}+\tfrac{(l-m+1)(l+m+1)}{(2l+1)(2l+3)}\left[(2\!-\!\tfrac{\al_{(0)}r_{(0)}}{2})\al_{(0)}r_{(0)} + (l\!-\!\al_{(0)}r_{(0)})\tfrac{r_{(0)}^2}{\ell^2}\right]\;.
\end{align}
Hardly prime-time programming, but it is what it is. Equating the $O(a^0)$ terms recovers Sfetsos's nonrotating series solution \cite{sfetsos}: 
\begin{equation}
f_{lm}^{(0)} = -\,\frac{l+\half}{u_{lm}^{(0)}}\,q_{(0)}\,\del_{m0}\;.
\end{equation}
Finally, and with only mild exuberance, I collect the $O(a^2)$ terms to present the leading rotational correction to the angular profile: 
\begin{equation}
f_{lm}^{(2)} = \frac{-1}{u_{lm}^{(0)}}\,\left[(l\!+\!\half)\,q_{(2)}\,\del_{m0}+ u_{lm}^{(2)}\,f_{lm}^{(0)} + v_{l-1,m}\,f_{l-1,m}^{(0)} + w_{l+1,m}\,f_{l+1,m}^{(0)} + x_{l-2,m}\,f_{l-2,m}^{(0)} + y_{l+2,m}\,f_{l+2,m}^{(0)} \right]\;.
\end{equation}
Goodnight moon. 
\section{Discussion}\label{sec:end}

I have calculated the gravitational backreaction from a massless particle on the future horizon of a Kerr-AdS black hole. This concludes, with bone-shattering finality, the gravity installment of my program for revitalizing black-hole statistical mechanics. 
\\\\
Next up: Chaos. 
\\\\
\begin{center}
	\textit{Acknowledgments}
\end{center}
I thank Joe Swearngin and Nick Hunter-Jones for preliminary work on this project and for feedback on the final product. I am grateful to Douglas Stanford for taking the time to read a draft and float potential applications of my formalism. I also thank Dante, Maria, and Sydney at the PI Bistro for moral support. This research was supported in part by Perimeter Institute for Theoretical Physics. Research at Perimeter Institute is supported by the Government of Canada through the Department of Innovation, Science and Economic Development Canada and by the Province of Ontario through the Ministry of Economic Development, Job Creation and Trade.

\bibliographystyle{unsrt}
\bibliography{References_2019-03-17}

\begin{thebibliography}{10}

\bibitem{shenker_stanford}
S.~H. Shenker and D.~Stanford.
\newblock Black holes and the butterfly effect.
\newblock {\em Journal of High Energy Physics}, 2014(3), 2014.

\bibitem{kitaev2018}
A.~Kitaev and S.~J. Suh.
\newblock The soft mode in the {S}achdev-{Y}e-{K}itaev model and its gravity
  dual.
\newblock {\em Journal of High Energy Physics}, 2018(5):183, 2018.

\bibitem{polchinski2017a}
J.~Polchinski.
\newblock Memories of a theoretical physicist.
\newblock {\em arXiv:1708.09093}, 2017.

\bibitem{bentov_swearngin}
Y.~BenTov and J.~Swearngin.
\newblock Gravitational shockwaves on rotating black holes.
\newblock {\em General Relativity and Gravitation}, 51(2), 2019.

\bibitem{dray_thooft}
T.~Dray and G.~'t~Hooft.
\newblock The gravitational shock wave of a massless particle.
\newblock {\em Nuclear Physics B}, 253:173--188, 1985.

\bibitem{gibbonsperrypope}
G.~W. Gibbons, M.~J. Perry, and C.~N. Pope.
\newblock The first law of thermodynamics for {Kerr}-anti-de {S}itter black
  holes.
\newblock {\em Classical and Quantum Gravity}, 22(9):1503--1526, 2005.

\bibitem{hawking_hunter_taylor-robinson}
S.~W. Hawking, C.~J. Hunter, and M.~M. Taylor-Robinson.
\newblock Rotation and the {A}d{S}-{CFT} correspondence.
\newblock {\em Physical Review D}, 59(6), 1999.

\bibitem{ghp}
R.~Geroch, A.~Held, and R.~Penrose.
\newblock A space-time calculus based on pairs of null directions.
\newblock {\em Journal of Mathematical Physics}, 14(7):874, 1973.

\bibitem{chand}
S.~Chandrasekhar.
\newblock {\em The mathematical theory of black holes}.
\newblock Clarendon Press, Oxford, 2009.

\bibitem{fels1989}
M.~Fels and A.~Held.
\newblock Kerr-{S}child rides again.
\newblock {\em General relativity and gravitation}, 21(1):61--68, 1989.

\bibitem{taub1981}
A.~H. Taub.
\newblock Generalised {K}err-{S}child space-times.
\newblock {\em Annals of Physics}, 134(2):326--372, 1981.

\bibitem{sfetsos}
K.~Sfetsos.
\newblock On gravitational shock waves in curved spacetimes.
\newblock {\em Nuclear Physics B}, 436(3):721--745, 1995.

\bibitem{maldacena2016b}
J.~Maldacena and D.~Stanford.
\newblock Remarks on the {S}achdev-{Y}e-{K}itaev model.
\newblock {\em Physical Review D}, 94(10), 2016.

\end{thebibliography}
\end{document}